\documentclass{PoS}

\title{General relativity experiment with frozen spin rings}

\ShortTitle{General relativity experiment with frozen spin rings}

\author{\speaker{Andr\'as L\'aszl\'o}\\
        Wigner Research Centre for Physics of the Hungarian Academy of Sciences,\\
        Konkoly-Thege M. u. 29-33, 1121 Budapest, Hungary\\
        E-mail: \email{laszlo.andras@wigner.mta.hu}}

\abstract{In experimental proposals published in the last two decades, 
a so called frozen spin storage ring concept was proposed for setting upper 
experimental bounds to electric dipole moment (EDM) of elementary particles. 
In a recent paper of ours, a fully covariant general relativistic (GR) 
calculation was presented on the Earth's gravitational modification 
effect in such mixed magnetic-electric frozen spin storage ring on the spin 
transport, which could contribute to such measurement. 
It was shown that similarly to an EDM signal, GR causes 
a spin precession out of the orbital plane, under the frozen spin condition. 
The rate of the vertical polarization buildup is predicted to be 
${-}a\,\beta\gamma\;\mathbf{g}/c$, where $\mathbf{g}$ is the gravitational acceleration on 
the surface of the Earth, $c$ is the speed of light, $\beta\gamma$ is the 
particle momentum over mass, and $a$ is its magnetic moment anomaly. 
It is seen that the effect increases unboundedly with the Lorentz factor 
$\gamma$. Moreover, it is proportional to the magnetic moment anomaly $a$. 
This paper mainly addresses the experimental perspectives to detect this 
effect in a realistic frozen spin storage ring configuration. Such a 
measurement would provide a novel test of GR, sampling the tensorial nature 
of GR at a microscopic level, as acting on the spin vector of 
elementary particles. The conclusion is that the pertinent GR experiment 
seems to be realistic with large magnetic moment anomaly particles, 
such as tritons, helion3 or protons, whereas it is not realistic with 
small magnetic moment anomaly particles, such as deuterons, muons or 
electrons.}

\FullConference{23rd International Spin Physics Symposium - SPIN2018 -\\
                10-14 September, 2018\\
                Ferrara, Italy}

\definecolor{persianred}{rgb}{0.8, 0.2, 0.2}
\definecolor{brownish}{rgb}{0.6, 0.2, 0.2}

\begin{document}

\section{Introduction}

In a recent set of papers \cite{morishima2018a,morishima2018b,morishima2018c} 
a claim was made by Morishima \emph{et al} that, in the muon anomalous magnetic 
moment experiments, also called $\mathsf{g}{-}2$ experiments \cite{g2,bennett2006,miller2007,mane2005}, 
there can be an unaccounted systematic error caused by the gravitational 
field of the Earth through its general relativistic (GR) action on spin 
vectors. Various GR authors \cite{visser2018,guzowski2018} responded with 
heuristic estimates, claiming that the effect must be way smaller. Further GR 
authors \cite{nikolic2018} argued that the effect exactly cancels. Also, 
due to the non-geodesic nature of the beam orbit, the usual formulae of 
de Sitter and Lense-Thirring precession \cite{lammerzahl2001} 
(also called geodetic effect) do not apply as they are. 
Motivated by the above controversy, a fully covariant GR calculation was 
performed by us in \cite{laszlo2018} for quantifying this effect in the 
$\mathsf{g}{-}2$ rings. It was shown that 
the final experimental systematics caused by GR is nonzero, but is indeed 
way too small to disturb the measurement. This smallness of the GR systematics 
is, however, a result of a not entirely trivial effect. First of all, the 
GR contribution is \emph{ab initio} small in comparison to the spin precession 
rate due to magnetic field in a $\mathsf{g}{-}2$ ring setting. 
Secondly, it turns out that the 
GR systematics comes as a beam-radial precession vector, which adds 
vectorially to the otherwise vertical magnetic precession vector caused by 
$\mathsf{g}{-}2$, thus making the GR modification in the observed 
precession frequency to be only of second order. In conclusion, the GR 
systematics in $\mathsf{g}{-}2$ rings caused by the gravitational field of 
the Earth is negligible.

As pointed out, the absence of substantial GR contribution by Earth's 
gravitational field to $\mathsf{g}{-}2$ experiments is partly a result of 
a second order suppression, caused by the geometry of the particular ring 
configuration. The pertinent GR modification effect, however, 
can be brought up to first order if the guiding fields of the particle 
beam is configured such that the spin vector in the comoving frame does 
not feel any torque by the guiding fields. 
Such a configuration is called a \emph{frozen spin storage ring} 
\cite{senichev2017,semertzidis2016,talman2017}, and that 
concept was mainly invented for experimental determination of electric 
dipole moment (EDM) \cite{talman2017} of charged particles with spin. 
Like since a decade ago, it was realized by other authors that in a 
frozen spin ring the Earth's gravity is likely to have a systematic contribution 
\cite{kobach2016,silenko2007,obukov2016a,obukov2016b,orlov2012}. 
These approximative calculations captured well the qualitative 
behavior of the GR contribution, and indicated that if the design sensitivity 
of EDM rings intends to reach ${\approx}10^{-29}\,e\mathrm{cm}$, the GR effect needs to be considered. 
In order to quantitatively capture the magnitude of the effect accurately, 
a fully covariant GR calculation was performed by us in \cite{laszlo2018}. 
The calculation shows that in a frozen spin configuration, Earth's 
gravitational field causes an out of the orbital plane precession at a rate
\begin{eqnarray}
 {-}a\;\beta\gamma\!\!\!\!\underbrace{\mathbf{g}/c}_{{\approx}33\,\mathrm{nrad/sec}}
\label{eqGRprec}
\end{eqnarray}
where $\mathbf{g}$ denotes the gravitational acceleration at the surface of 
the Earth, $c$ is the speed of light, $a$ is the magnetic moment anomaly 
of the particle, and $\beta\gamma$ is the momentum over mass of the particle. 
It is seen that the effect is proportional to the magnetic moment anomaly $a$, 
and that it grows unboundedly with the Lorentz factor $\gamma$. Since the 
effect is in principle unbounded, the question naturally arises: can one 
specifically design an experiment such that this GR contribution is well 
measurable? One could think, for instance, about optimizing EDM rings such 
that this effect becomes dominant. If such an experiment could be 
carried out, it would test GR in a very interesting condition:
\begin{itemize}
 \item with microscopic particles,
 \item at relativistic speeds,
 \item along non-geodesic (forced) trajectories,
 \item tensorial nature of GR would be at test and not merely the gravitational drag.
\end{itemize}
Such a configuration would be a microscopic analogy of the Gravity Probe B 
satellite experiment \cite{grav,everitt2011}, where GR modification of 
free gyroscopic motion around Earth was measured. However, in contrast to 
satellite experiments, in frozen spin rings the general relativistic 
gyroscopic motion would be tested along non-geodesic (forced) trajectories. 
That is, the GR correction to the \emph{Thomas precession} could be 
experimentally tested.

\section{Relativistic motion of charged point particle with spin, in GR}

\begin{figure}[!h]
\begin{center}
\begin{minipage}{12cm}
{\color{blue}$u^{a}$ denotes the four velocity of the particle at points of the trajectory}\\
\\
{\color{brownish}$w^{a}$ denotes the spin direction four vector at points of the trajectory}\\
\\
{(one has the orthogonality constraint $g_{ab}\,u^{a}\,w^{b}=0$ at each point)}
\end{minipage}\begin{minipage}{1.5cm}\includegraphics[width=1.25cm]{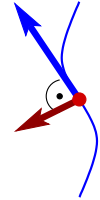}\end{minipage}
\end{center}
\caption{Illustration of the spacetime evolution of a point particle with spin. 
The particle evolves along some future directed worldline with instantaneous 
four velocity $u^{a}$. At the points of the worldline a spin direction vector 
$w^{a}$ is also present. At each point of the worldline one has the constraints 
$\;g_{ab}\,u^{a}\,u^{b}=1$, $\;g_{ab}\,w^{a}\,w^{b}=-1\;$ and 
$\;g_{ab}\,u^{a}\,w^{b}=0\;$ where $\;g_{ab}\;$ is the spacetime metric 
tensor field with signature (+,-,-,-).}
\label{figpointp}
\end{figure}

In a special relativistic or general relativistic framework, the motion 
of a point particle, i.e.\ of a particle with irresolvable internal structure 
is encoded via a future directed timelike worldline in spacetime. At each 
point of its worldline, it has a corresponding tangent vector, which can 
be normalized to unit length, and is then called the four velocity $u^{a}$. 
Moreover, if the particle has a spin, then at each point of the worldline, 
a unit length spacelike vector $w^{a}$ describes the instantaneous spin 
direction, being always orthogonal to the four velocity $u^{a}$ due to its 
quantum mechanical origin. This is illustrated in Figure~\ref{figpointp}.

The evolution of the trajectory is governed by the relativistic Newton equation, 
and along the worldline the evolution of the spin direction vector is governed 
by the Thomas-Bargmann-Michel-Telegdi (TBMT) equations \cite{conte1996,jackson1999,hawking1973,wald1984}:
\begin{eqnarray}
 (A)\quad u^{a}\,\nabla_{a}u^{b} & = & -\frac{q}{m}\,g^{ab}\,F_{ac}\,u^{c}, \qquad\qquad\qquad\qquad\qquad\qquad\qquad(\leftarrow\,\mathrm{relativistic\;Newton\;eq.}) \cr
 & & \cr
 (B)\quad\;\;\,\, D^{F}_{u}w^{b}         & = & -\frac{\mu}{s}\, \left(g^{ab}\,F_{ac} \,-\, u^{b}\,u^{d}\,F_{dc} \,-\, g^{ab}\,F_{ad}\,u^{d}\,u^{e}\,g_{ec}\right) \,w^{c}, \quad(\leftarrow\,\mathrm{TBMT\;eq.})\cr
 & &                                   +\frac{d}{s}\, \left(g^{ab}\,{}^{\star}\!\!F_{ac} \,-\, u^{b}\,u^{d}\,{}^{\star}\!\!F_{dc} \,-\, g^{ab}\,{}^{\star}\!\!F_{ad}\,u^{d}\,u^{e}\,g_{ec}\right) \,w^{c}.
\label{eqnt}
\end{eqnarray}
\newpage

\noindent
In the above equations $m$, $q$, $s$ denote particle mass, charge 
and spin magnitude ($s=\frac{1}{2},1,\frac{3}{2},\dots$), whereas $\mu$ and 
$d$ denote the magnetic and electric dipole moment of the particle, respectively. 
The electric dipole moment $d$ is usually extremely small, and the aim of the 
EDM experiments is to obtain an upper bound to the quantity $d$. The field 
quantities in the above equations are as follows: $F_{bc}$ denotes the electromagnetic 
field strength tensor of the guiding fields, ${}^{\star}\!\!F_{bc}$ denotes the 
Hodge dual of the electromagnetic field strength tensor, $g_{ab}$ denotes the 
spacetime metric tensor field as usual, $\nabla_{a}$ denotes the spacetime 
covariant derivation compatible with the metric, and $D^{F}_{u}w^{b}$ denotes 
the \emph{Fermi-Walker derivative} of $w^{b}$ along the worldline described 
by $u^{a}$, defined as 
$D^{F}_{u}w^{b}=u^{a}\,\nabla_{a}w^{b}+g_{ed}w^{e}u^{b}u^{a}\nabla_{a}u^{d}-g_{cd}w^{c}u^{d}u^{a}\nabla_{a}u^{b}$. 
As it is well-known, the geometric meaning of the covariant derivative of 
vectors is the infinitesimal parallel transport, whereas the meaning of Fermi-Walker 
derivative is the infinitesimal parallel transport of rigid orthonormal 
frames (preserving orthonormality). As such, the spin evolution equation for 
free gyroscopes would be $D^{F}_{u}w^{b}=0$, and the Gravity Probe B satellite 
experiment \cite{grav,everitt2011} tested this kinematic equation. Whenever 
one has geodesic motion ($u^{a}\nabla_{a}u^{b}=0$), the kinematic effect of 
the gyroscopic motion is called geodetic effect (de Sitter or Lense-Thirring precession). Along 
a forced trajectory ($u^{a}\nabla_{a}u^{b}=\mathrm{some\;four\;force}$), the 
kinematic effect of the gyroscopic motion is called \emph{Thomas precession}, 
which is the case for particles in Earth-based storage rings. Moreover, the 
explicit torque exerted on the spin by the electromagnetic field, i.e.\ by the 
right hand side of Eq.(\ref{eqnt})(B), is called \emph{Larmor precession}. The 
Thomas and Larmor precession is already present in the special relativistic 
(flat spacetime) case. The frozen spin condition means that assuming the 
absence of gravity and of EDM, one tries to construct a configuration in 
which the Thomas+Larmor precession vanishes.

\section{Frozen spin (EDM) rings and GR modification of spin dynamics within}

\begin{figure}[!h]
\begin{center}
\begin{minipage}{\textwidth}
\begin{center}
\begin{minipage}{10.5cm}\includegraphics[width=10cm]{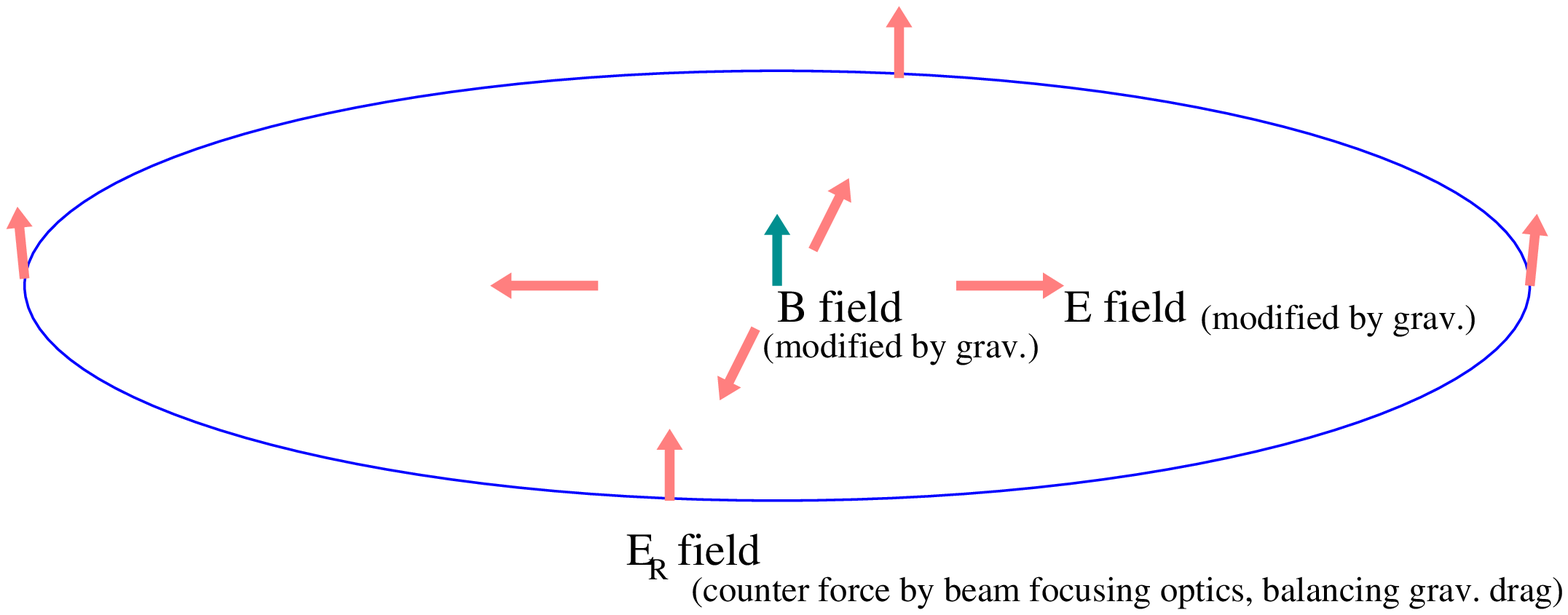}
\end{minipage}\begin{minipage}{3.4cm}\vspace*{-16mm}\includegraphics[width=2.8cm]{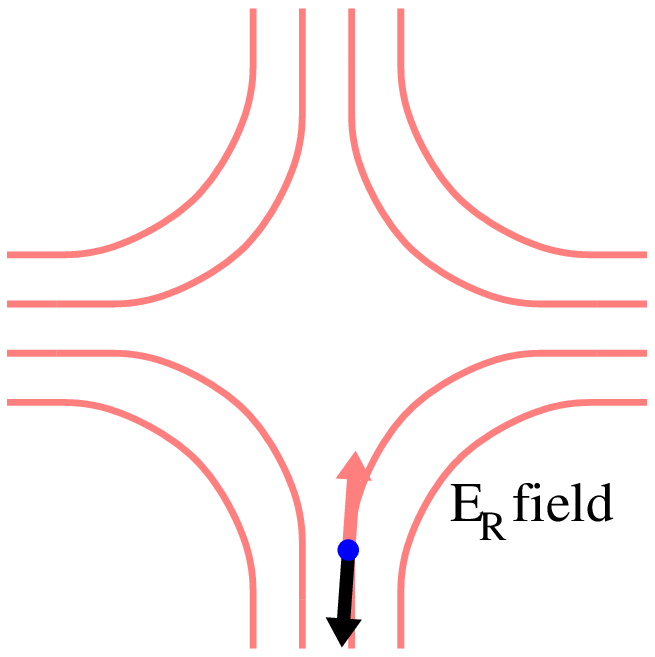}\newline{\hspace*{0.2cm}\tiny(grav.drag balanced by vert.focusing)}\end{minipage}
\end{center}
\end{minipage}
\end{center}
\caption{Illustration of the electromagnetic fields in a frozen spin storage 
ring, with combined magnetostatic and electrostatic bending fields 
($B$ and $E$), and with electrostatic vertical focusing beam optics. 
The geometry of the bending fields need to be defined over the curved spacetime 
due to the gravitational field of the Earth. Moreover, 
Earth's gravitational drag is equilibrated by the Earth-radial 
electrostatic field ($E_{R}$), exerted by the electrostatic 
vertical beam focusing optics at the place of the nominal 
equilibrium beamline.}
\label{figfields}
\end{figure}

The main idea of frozen spin storage rings \cite{senichev2017,semertzidis2016,talman2017} 
is that in an idealized case one injects longitudinally spin polarized particle 
beams into a planar circular storage ring, where the magnetic bending field 
($B$) is homogeneous vertical, and there is a horizontal cylindric, i.e.\ beam-radial 
electric bending field ($E$) in addition. Given the particle type, the bending 
fields $B$ and $E$ are configured such that both the relativistic Newton 
equation Eq.(\ref{eqnt})(A) is satisfied, moreover the particle spin, which 
is governed by the TBMT equation Eq.(\ref{eqnt})(B), does not precess against 
the beam-tangential direction in the horizontal plane (\emph{horizontal frozen spin condition}). 
If no EDM of the particle and no gravitational fields are present, then the 
spin would not precess at all in such a configuration against the beam-tangential 
direction (\emph{frozen spin condition}). Whenever an EDM of the particle would 
be present, the spin would precess out of the orbital plane under this condition. 
Similarly, the gravitational field of the Earth through GR causes an out of the 
orbital plane precession, which was qualitatively determined in 
\cite{kobach2016,silenko2007,obukov2016a,obukov2016b}, and quantitatively 
determined in \cite{laszlo2018,orlov2012}. Given the particle mass $m$, charge 
$q$, spin $s$, gyromagnetic factor $\mathsf{g}=\frac{2\,m\,\mu}{q\,s}$, magnetic 
moment anomaly $a=\frac{\mathsf{g}-2}{2}$, as well as its momentum over mass 
$\beta\gamma$, and ring radius $r$, the Newton equation for the 
circular motion and the frozen spin condition are simultaneously satisfied whenever
\begin{eqnarray}
E\cdot r & \;=\; & -\mathrm{sign}(a)\,\frac{m\,c^{2}}{q}\,\frac{(a\,\beta\gamma)^{2}\sqrt{a^{2}+(a\,\beta\gamma)^{2}}}{a^{2}\left(1+a\right)}, \cr
B\cdot r     & \;=\; & \qquad\qquad\;\, \frac{m\,c}{q}\;\;\frac{(a\,\beta\gamma)(a-(a\,\beta\gamma)^{2})}{a^{2}\left(1+a\right)}
\label{eqfields}
\end{eqnarray}
holds (two equations for the two unknowns $B$ and $E$). Under this condition, 
assuming negligible EDM and GR effects, the longitudinal spin polarization of the 
beam is preserved, i.e.\ no vertical polarization buildup shall be present. 
If a small nonvanishing particle EDM is present, that would cause a slow 
vertical spin polarization buildup, which the EDM experiments would aim to 
detect. On the other hand, assuming negligible EDM, but non-negligible GR 
effect by Earth's gravitational field, the following ingredients are affected:
\begin{itemize}
 \item The metric and thus the parallel transport $\nabla_{a}$ and Fermi-Walker transport $D^{F}$ are modified.\newline
       (The Newton + TBMT equations Eq.(\ref{eqnt}) are affected by GR, quite naturally.)
 \item The Maxwell equations $\;g^{ab}\,\nabla_{a}F_{bc}=0$, $\;g^{ab}\,\nabla_{a}{}^{\star}\!\!F_{bc}=0\;$ are modified.\newline
       (The electromagnetic fields of the storage ring are affected by GR, as illustrated in Figure~\ref{figfields}.)
\end{itemize}

In \cite{laszlo2018} it is shown that the total GR effect induces 
an out of the orbital plane precession at a rate Eq.(\ref{eqGRprec}) in a 
generic mixed magnetic-electric frozen spin ring with electric vertical focusing. 
An earlier calculation \cite{orlov2012} quantified the same thing for the 
special case of electrostatic-only frozen spin rings 
(only possible for $a>0$ and at \emph{magic momentum} $|\beta\gamma|=\frac{1}{\sqrt{a}}$). 
It can be shown that the GR prediction differs from the prediction 
of semi-classical treatment of gravity at a ratio $a:(1+a)$, and thus the 
experimental verification of the effect would be truly a GR test. 
(Semi-classical treatment: one prescribes by hand the four-force of the 
gravitational drag on the particle as coming from GR, i.e.\ as coming from 
the gravitational acceleration, but one does the calculation over flat 
spacetime.)

\section{Optimalization}

From Eq.(\ref{eqGRprec}) it is seen that one can increase the GR effect via 
trying to reach as large $\left\vert a\,\beta\gamma\right\vert$ as possible. 
However, from Eq.(\ref{eqfields}) it is seen that for a given particle type 
(with fixed $\frac{m}{q}$ and $a$), the necessary electric bending power 
$\left\vert E\cdot r\right\vert$ for maintaining the frozen spin condition 
increases rapidly, as ${\sim}O\big(\left\vert a\,\beta\gamma\right\vert^{3}\big)$. 
Since too large ring radius $r$ or too large electric fields $\left\vert E\right\vert$ 
are unrealistic, this sets limits to the experimental possibilities. 
One can realize, however, that because of Eq.(\ref{eqfields}), for a fixed 
$\left\vert a\,\beta\gamma\right\vert$, the necessary electric bending power 
$\left\vert E\cdot r\right\vert$ decreases as ${\sim}O(\left\vert a\right\vert^{-2})$. 
Thus, it is advantageous to use particles with large magnetic moment anomaly $\left\vert a\right\vert$. 
Using particle data table of \cite{stone2005}, a set of particle species 
can be proposed for conducting such a frozen spin GR experiment. Assume that 
a reasonable GR signal strength of $\left\vert a\,\beta\gamma\right\vert=0.4$ 
is aimed to be achieved, which would result in an out of the orbital plane 
precession (vertical polarization buildup) due to GR at a rate 
$13.1\,\mathrm{nrad/sec}$, according to Eq.(\ref{eqGRprec}). Assume that 
the main bottleneck of feasibility, namely the electric field strength, is 
also fixed to a reasonable value $|E|=4.10\,\mathrm{MV/m}$. Then, 
feasible and unfeasible configurations are listed in Table~\ref{tablerealistic}.

\begin{table}[!h]
\textit{Particle species with which frozen spin GR experiments may be conducted:}
\begin{center}
\begin{tabular}{c|c|c|c|c|c|c}
 particle name & $\left\vert a\,\beta\gamma\right\vert$ & $r$ $\mathrm{[m]}$ & $\left\vert E\right\vert$ $\mathrm{[MV/m]}$ & $\left\vert B\right\vert$ $\mathrm{[Tesla]}$ & $p$ $\mathrm{[MeV/c]}$ & $\mathcal{E}_{\mathrm{kin}}$ $\mathrm{[MeV]}$ \cr
\hline
 triton        & $0.4$                       & $1.55$             & $4.10$                               & $0.0335$                          & $141.9$               & $3.58$ \cr
 helion3       & $0.4$                       & $4.13$             & $4.10$                               & $0.0353$                          & $268.5$               & $12.8$ \cr
 proton        & $0.4$                       & $7.50$             & $4.10$                               & $0.0304$                          & $209.7$               & $23.1$ \cr
\end{tabular}
\end{center}

\vspace*{2mm}
\textit{Particle species with which frozen spin GR experiments are unrealisticly expensive:}
\begin{center}
\begin{tabular}{c|c|c|c|c|c|c}
 particle name & $\left\vert a\,\beta\gamma\right\vert$ & $r$ $\mathrm{[m]}$ & $\left\vert E\right\vert$ $\mathrm{[MV/m]}$ & $\left\vert B\right\vert$ $\mathrm{[Tesla]}$ & $p$ $\mathrm{[MeV/c]}$ & $\mathcal{E}_{\mathrm{kin}}$ $\mathrm{[MeV]}$ \cr
\hline
 deuteron      & $0.4$                       & $1796$             & $4.10$                               & $0.0243$                         & $5283$               & $3731$ \cr
 electron      & $0.4$                       & $5942$             & $4.10$                               & $0.0136$                         & $176.5$              & $176.0$ \cr
 muon          & $0.4$                       & $1228520$          & $4.10$                               & $0.0136$                         & $36497$              & $36391$ \cr
\end{tabular}
\end{center}
\caption{Top table: list of particle species with which realistic frozen spin 
GR experiments can be conducted. Some realistic example configurations 
are also listed along with the particle types. Namely, the predicted GR 
signal strength is fixed to $\left\vert a\,\beta\gamma\right\vert=0.4$, meaning 
an expected vertical polarization buildup rate of $13.1\,\mathrm{nrad/sec}$. At the same 
time the electric bending field, being the main bottleneck, 
is fixed to a technically realistic value $|E|=4.10\,\mathrm{MV/m}$. 
Bottom table: list of particle species with which frozen spin GR experiments 
are unrealistically expensive. Those particles all have small magnetic moment 
anomaly $\left\vert a\right\vert$.}
\label{tablerealistic}
\end{table}

The biggest systematic errors in such an experimental setting is expected 
to come from the \emph{conical imperfections} of the magnetic bending field. 
That is, besides the vertical component $B$, there can be a small beam-radial 
magnetic field component $B_{r}$. Assuming a planar circular beam orbit at 
equilibrium on the Earth's surface satisfying the horizontal frozen spin 
condition, then its vertical polarization buildup rate is predicted to be
\begin{eqnarray}
 \underbrace{-\frac{q\,(1+a)}{m}\frac{1}{\gamma^{2}}\,B_{r}}_{\mathrm{magnetic\;field\;conicality\;imperfection\;term}} \quad+\quad \underbrace{-a\,\beta\gamma\;\mathbf{g}/c}_{\mathrm{GR\;term}}
\label{eqconical}
\end{eqnarray}
due to magnetic field conical imperfection and GR. The magnetic field conical 
imperfection $\frac{B_{r}}{B}$ is normally rather hard to suppress beyond 
certain limits. In order to overcome that limitation, in \cite{talman2018} a 
new concept was proposed: two different particle species are proposed to be 
studied in the same frozen spin storage ring, called to be a \emph{doubly-frozen spin ring}. 
That turns out to be kinematically possible for certain pairs of particle 
species, at unique beam momenta and bending field configurations. Given such a 
setting, one has two vertical polarization buildup observables 
Eq.(\ref{eqconical}) for the two particle beams, and one can use these to 
create an optimal linear combination of these observables which is insensitive 
to $B_{r}$ to the first order. In that way, the extreme sensitivity to the 
magnetic field imperfections can be suppressed.

\section{Concluding remarks}

In this contribution a general relativity (GR) experiment was proposed in 
order to measure the effect of Earth's gravitational field on the spin 
transport in frozen spin particle storage rings. Since electric dipole moment 
(EDM) experiments are already planned to be performed using frozen spin rings 
\cite{senichev2017,semertzidis2016,talman2017}, it is quite natural to 
try to measure this effect as part of their program. GR causes a precession 
out of the orbital plane at a rate $\;{-}a\,\beta\gamma\;\mathbf{g}/c\;$ \cite{laszlo2018}, where $\;\mathbf{g}\;$ is the gravitational 
acceleration at the surface of the Earth, $\;c\;$ is speed of light, $\;a\;$ is 
the magnetic moment anomaly of the particle, and $\;\beta\gamma\;$ is the momentum 
over mass of the particle. The constant 
$\;\mathbf{g}/c\;$ is about $\;{\approx}33\,\mathrm{nrad/sec}$. Since the effect is 
proportional to the product $a\,\beta\gamma$, it is advantageous to have 
as large $\left\vert a\,\beta\gamma\right\vert$ as possible. The issue is 
caused by the fact that the necessary electric bending field increases 
very fast, as $\;{\sim}O\big(\left\vert a\,\beta\gamma\right\vert^{3}\big)$ 
for a fixed particle type. 
However, at a fixed anticipated GR signal strength, the necessary 
electric bending field decreases as $\;{\sim}O\big(\left\vert a\right\vert^{-2}\big)$, 
thus it is advantageous to use particles with large $\left\vert a\right\vert$. 
A relatively large GR signal strength ${\approx}10\,\mathrm{nrad/sec}$ can be met with a realistic bending radius 
${\approx}10\,\mathrm{m}$ and a realistic electric bending field 
${\approx}5\,\mathrm{MV/m}$ with moderate energy 
triton, helion3 or proton beams, due to their large magnetic moment anomaly 
$\left\vert a\right\vert$. Particles with small $\left\vert a\right\vert$ such 
as deuterons, muons or electrons are not realistic options. The main 
systematics, caused by magnetic bending field imperfections can be 
substantially reduced via the method of doubly-frozen spin ring concept 
\cite{talman2018} when in the same bending fields two beams with different 
particle species are compared. This opens up an experimental possibility to 
test GR in a very interesting regime: with microscopic particles, at 
relativistic speeds, along non-geodesic (forced) trajectories, and to test 
the tensorial nature of GR under these conditions --- not merely the 
gravitational drag. The pertinent GR signal is distinguishable from a possible 
EDM signal via their opposite space reflection properties, i.e.\ via their 
opposite sign change when reverting beam direction.

\section*{Acknowledgments}

This work was supported in part by the Hungarian Scientific Research fund 
(NKFIH 123842-123959).

\end{document}